# Design and Validation of a Modular Smart Headband with Embroidered Electrodes for Comfortable EEG Monitoring

Komal K. [1,2,3]*, Cleary F.[2], Narayanan R. P.[2], Wells J S.G.[1], Buiatti M.[3a], and Bennett L.[1a]

[1] School of Health Sciences, Southeast Technological University, Cork Road, Waterford, Ireland

[2] Walton Institute, Southeast Technological University, Cork Road, Waterford, Ireland

[3] CIMeC, Center for Mind/Brain Sciences, University of Trento, Rovereto, Italy.

[a]Equally contributed as senior authors.

*Abstract*— **The wearable EEG device sector is advancing rapidly, enabling fast and reliable detection of brain activity for investigating brain function and pathology. However, many current EEG systems remain challenging for users with neurological conditions due to bulky wiring, lengthy skin preparation, gel-induced discomfort, risk of irritation, and high cost, all of which limit long-term monitoring. This study presents a proof-of-concept smart modular headband incorporating adjustable, replaceable embroidered electrodes for EEG acquisition. Compared with conventional devices, the smart headband reduces wiring complexity, removes the need for skin preparation, and minimizes irritation associated with gel-based electrodes. Its modular structure allows adjustable fitting without requiring multiple size options, enhancing comfort and adaptability for everyday EEG monitoring.**

**The smart headband prototype was tested on 10 healthy university students using three behavioral tasks: (1) eyes open/closed, (2) auditory oddball, and (3) visual oddball paradigms. The smart headband successfully captured alpha peaks during the eyes-open/closed task ($p = 0.01$) and reliably recorded the event-related potentials associated with the oddball effects - the auditory P300 ($p = 0.014$) and the visual N170 ($p = 0.013$) - demonstrating an equivalent performance to a commercial sponge-based EEG cap. A user survey indicated improved comfort and usability, with participants reporting that the soft, structurally designed headband enhanced wearability relative to a conventional cap. Overall, this prototype provides a comfortable, modular, and cost-effective solution to reliable EEG monitoring in real-world applications.**

*Index Terms*—**Smart headband design, 3D embroidered electrodes, EEG measurement.**

## I. INTRODUCTION

ELECTROENCEPHALOGRAPHY (EEG) is a non-invasive technique for investigating brain function and pathology by recording the electrical activity of neural origin from the scalp [1]. Thanks to recent technological advances, wearable EEG devices present novel opportunities for product development and therapeutic solutions for pathological conditions [2]. Given the non-invasive, real-time monitoring of EEG signals through wearable technology, several studies utilise

EEG data features and observable patterns to interpret brain signals, thereby enabling the representation of neural activity evoked by external stimuli [3], [4], [5], [6], [7], [8]. However, current wearable EEG acquisition systems often involve numerous wires, multiple gel-based or metallic electrodes, and bulky amplifiers. The process of electrode placement and scalp preparation is time-consuming, and prolonged use of such complex devices for extended monitoring has been reported as uncomfortable and impractical for the person [9]. Furthermore, unstable contact between the scalp and the electrodes increases impedance and affects signal quality. To achieve a reliable signal, additional pressure is often required to improve electrode contact; however, applying this pressure can contribute to user discomfort [10].

Clinical EEG systems currently in use still rely on medical-grade Ag/AgCl electrodes, which detect electrical activity from the scalp surface [11]. The application of these electrodes requires an electrolyte gel to make contact with the scalp, reducing skin–electrode impedance and providing mechanical stability. However, these electrodes necessitate time-consuming skin preparation, and the electrolyte gel tends to dry out over time, increasing susceptibility to interference, signal degradation, and a reduced signal-to-noise ratio [12], [13]. Additionally, prolonged use of gel-based electrodes can cause skin irritation, such as rashes and allergic reactions [14]. Recent advancements in dry metallic disc electrodes have addressed some limitations of gel-based electrodes. Dry electrodes are easier to apply, eliminate the need for gel, and enhance overall system usability [14]. Nonetheless, they typically exhibit higher electrode–scalp impedance than gel-based counterparts [15]. Moreover, their rigid structure is often uncomfortable for the participants, and it can result in an unstable contact with the scalp, increasing the risk of motion artifacts and noise, which may degrade signal quality [16], [17].The development of a sponge-based commercial EEG cap, which uses soft sponges soaked in a saline solution, has further improved usability [18]. These caps enable faster preparation compared to traditional headsets while delivering equivalent or superior signal quality. They also offer improved comfort and make it easier to adjust electrode impedance for accurate EEG measurement. However, they are expensive, and they still depend on a wired amplifier, which limits individual mobility and can be inconvenient due to managing and carrying cable connections [18].

In recent years, various textile-based electrodes have been developed using soft materials such as polymers, graphene, conductive inks, mask deposition, laser printing, smart conductive fabrics, and embroidered techniques [19], [20], [21], [22], [23], [24], [25]. The key advantages of textile-based electrodes are that they record electrical activity without the need for any contact material (gel, saline solution), are very comfortable to wear, and are generally very cheap. However, there is still a demand for reliable textile-based EEG technologies that can enable long-term monitoring while prioritising user comfort and improving usability [10], [26]. Also, initial attempts have been made to compare their signal quality with medical-grade electrodes [17]. Most studies lack comprehensive validation of textile-based electrodes for real-world applications, such as using them to assess cognitive brain function [25], [26].

This paper presents the design and development of an innovative low-cost modular smart headband with seven integrated embroidered electrodes for EEG recording. The proposed smart headband improves user comfort, and its modular design can be adjusted to fit any head size, reducing the need to purchase multiple commercial EEG caps. It is easy to wear and eliminates the bulkiness caused by the multiple wires and carrying amplifier used in traditional EEG caps. Additionally, integrating an embedded Enbio8 controller makes the system portable and enables wireless transmission of EEG data. Moreover, no skin preparation is required for wearing the smart headband, reducing the skin preparation time and also eliminating the risk of skin irritation or allergic reactions. Ultimately, wearing a smart headband improves usability, wearability, and comfort, making long-term monitoring of EEG measurements more practical for the participants.

As a second key step, this study addresses the gap in the validation of embroidered electrodes by testing the performance of the presented headband in assessing cognitive brain function and comfort level. The performance of the smart headband was validated by: (1) testing its efficacy in obtaining reliable neural responses during eyes open/eyes closed rest and classical cognitive tasks (visual and auditory oddball) in comparison with a commercial sponge-based wet electrode EEG cap; and (2) assessing its comfort, wearability, and usability relative to the commercial EEG cap.

## II. DESIGN & IMPLEMENTATION

This section describes the fabrication of an embroidered electrode, the design and creation of the smart headband, and the hardware setup used for EEG measurement.

### *A. Fabrication of embroidered electrode:*

A Textile-based 3D embroidered electrode was designed and fabricated using the ZSK technical embroidery machine, incorporating EPC Win software [28]. The design image was initially created using the EPC Win software, where parameters such as shape, stitch pattern, stitch length, and stitch distance were selected. Subsequently, the digitalised design image was then utilised to fabricate the electrode using the ZSK technical embroidery machine. Previously, a series of samples with variations in structural design parameters was utilized to optimize the performance of textile electrodes. Electrical impedance analysis, mechanical characterization, including washing, and strength testing for failure analysis were conducted to nominate the best sample design [29]. Based on those characterization results, the structural design was further modified and utilized in the present research work. A circular shape, 20 mm in diameter, cross-stitch stitch patterns, stitch length of 2.2mm, and stitch distance of 2.5mm were selected.

The software design of the 3D embroidered electrode consists of three stitch layers: (1) a boundary layer to define the shape of the electrodes; (2) a stabilizing middle layer with non-compact baseline stitches; and (3) a final stitch layer of conductive electrode stitches for biopotential measurements in the scalp. The fabrication of a multi-layered 3D embroidered electrode is presented in Fig. 1.

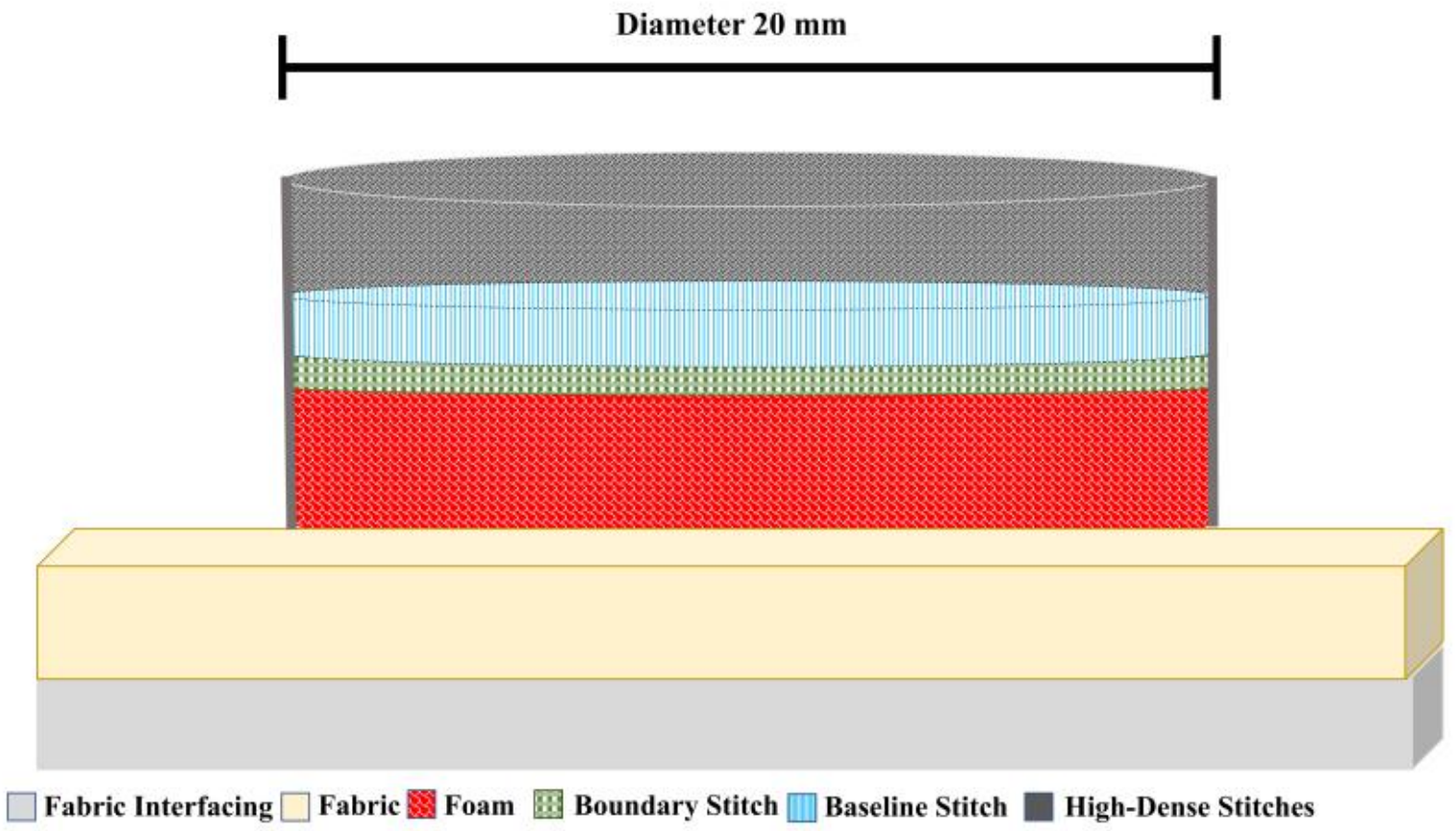

**Fig. 1.** Multi-layered 3D embroidered electrode

The fabrication process of the 3D embroidered electrode comprises three distinct layers: (1) a calico 100% cotton baseline fabric along with an interfacing layer; (2) a 2mm thick foam layer (commonly used for crafting due to its ease of cutting and adapting to customised shapes of the electrode [30] to raise the 3D effect to enhance scalp-electrode contact for accurate EEG measurements); and (3) the conductive stitch sub-layers for the measurement of biopotentials [31]. The Madeira HC-12 conductive thread was employed for the fabrication of embroidered electrodes, whose resistivity is less than 100 ohms per meter [32].

*B. Creation of a Smart headband modular design*

The smart headband was constructed using a neoprene sports band with dimensions of 60 cm long, 8.5 cm wide at the edges, 5 cm wide at the center, with a thickness of 3 mm of the fabric, and a weight of approximately 20 g, suitable for most adult head sizes [33]. This neoprene was selected due to its excellent fixation properties and its ability to hold hair in place during movement, ensuring stability and comfort during physical activities. Furthermore, this neoprene features an inner lining to enhance comfort, and a velcro fastener at the end allows for easy adjustment and removal.

A total of six 3D embroidered electrodes were integrated into the smart headband, positioned according to the 10–20 international electrode placement system at locations TP7, T7, FP1, FP2, T8, and TP8 [10], [34], [35], [36]. These positions were cross-checked against the series of tasks selected for EEG measurement in this study and verified with relevant literature [9], [10], [35]. Fig. 2 illustrates the step-by-step creation of the smart headband design. Within the modular headband design architecture, the basic module consists of a 3D embroidered electrode fabricated on a 100% calico cotton baseline fabric (see Fig. 2 (a)).

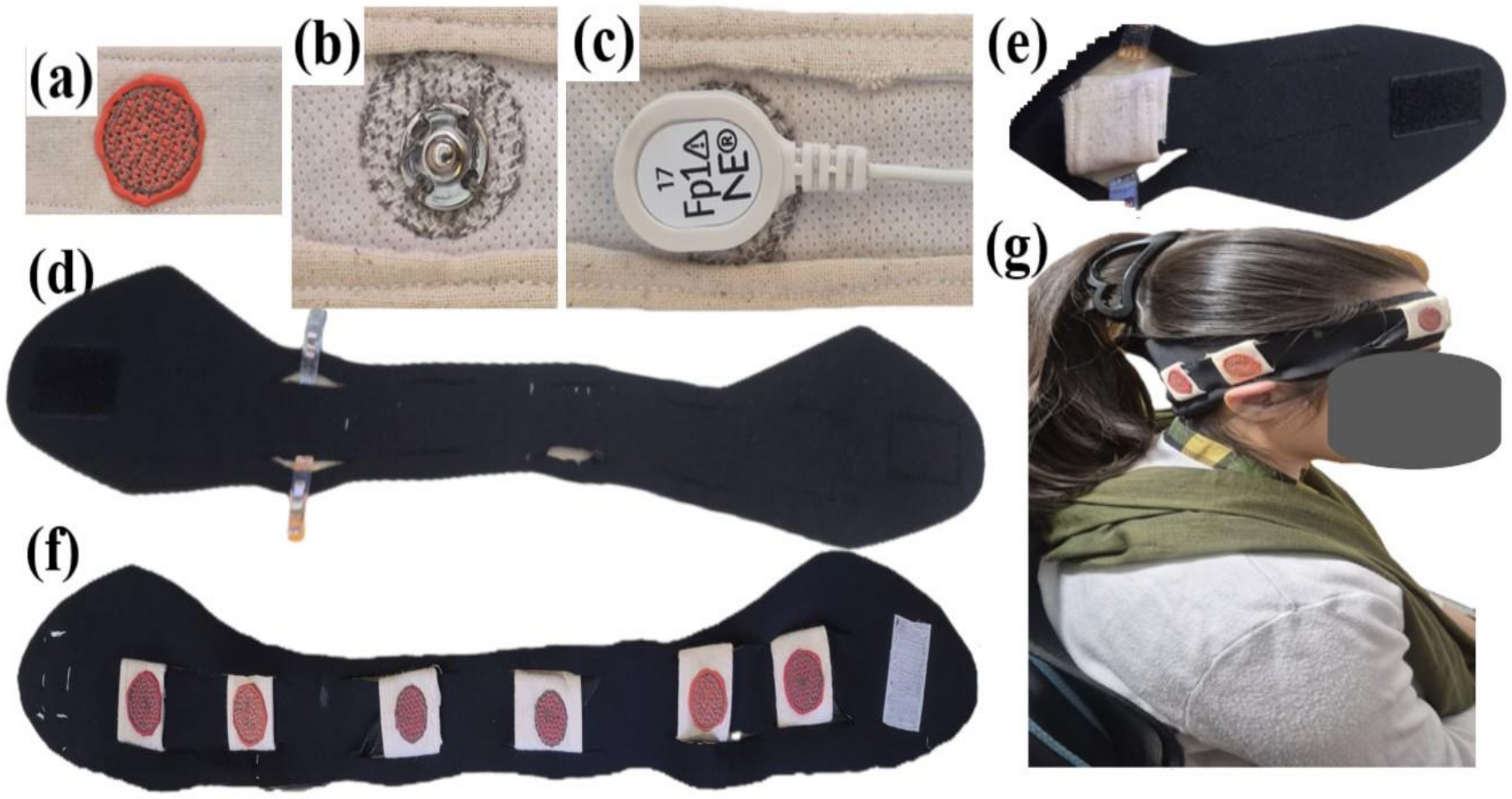

**Fig. 2.** Smart headband embedded with embroidered electrodes; (a) Embroidered electrode module (front view), (b) Back view (snap fastener for signal measurement); (c) Integration of the embroidered electrode with hardware wires; (d) Inclusion of twelve slits (six pairs) within neoprene headband for electrode placement; (e) Fixation of the electrode module within the slit; (f) Full headband integrated with embroidered electrodes; (f) Full headband integrated with embroidered electrodes; and (g) Smart headband worn on the subject.

Each fabric-integrated electrode module was cut to 100 × 50 mm, and its edges were hemmed with white non-conductive nylon thread. The electrode measured 20 mm in diameter, with an additional 5mm of surrounding calico fabric (in total 25mm). The center of the electrodes was sewn with a 13 mm conductive snap with HC-12 conductive Madeira thread on the backside of the electrode, enabling the connection to external wires. The conductive side of the electrode contacted the skin for EEG signal acquisition, while the conductive snap fastened on the back side allowed the direct connection to the Neuroelectrics system controller for EEG measurements.

Six such prepared modules were housed in a smart headband to accommodate lateral adjustment for different head sizes while maintaining a functional interface for EEG measurement. This modular design enables adaptability for individuals of varying sizes and facilitates easy electrode replacement. In order to enable the movement of the embroidered electrode module, the neoprene headband features twelve horizontal slits (six pairs) of 40 mm-long slits along its 60 cm length, allowing for electrode placement. These horizontal slits are positioned 1 cm from the top and bottom, respectively, leaving 3 cm vertical central space to house each embroidered electrode module (20 mm diameter of embroidered electrode surrounded by approximately 5mm calico fabric). Each 25 mm wide embroidered electrode module was housed within the slit and was secured with a velcro strip at its ends on the back side within the headband slit. This will allow for approximately 15 mm lateral adjustment for optimal positioning based on an individual's head size.

This embroidered electrode module is designed to be used to adjust electrodes at the 10–20 electrode system locations for different head sizes. Additionally, this velcro strip on the backside of the embroidered electrode modules allows individual modules to be removed for washing or replaced in case of fraying or damage, without needing to replace the entire smart headband. This modular smart headband design ensures comfort, adjustability, and modularity while maintaining appropriate electrode-skin contact. The reusable application of the smart headband features an embroidered electrode design, providing a low-cost alternative to traditional reusable gel-based electrodes. Additionally, the slit-based architectural design of the smart headband prototype provides flexibility to accommodate different head sizes, and it facilitates easy replacement during washing or in case of individual electrode degradation, thereby enhancing usability in practical EEG measurements.

### *C. EEG Measurement Setup*

The smart headband EEG measurement setup involved various steps from electrode placement to recording EEG data, as illustrated in Fig. 3.

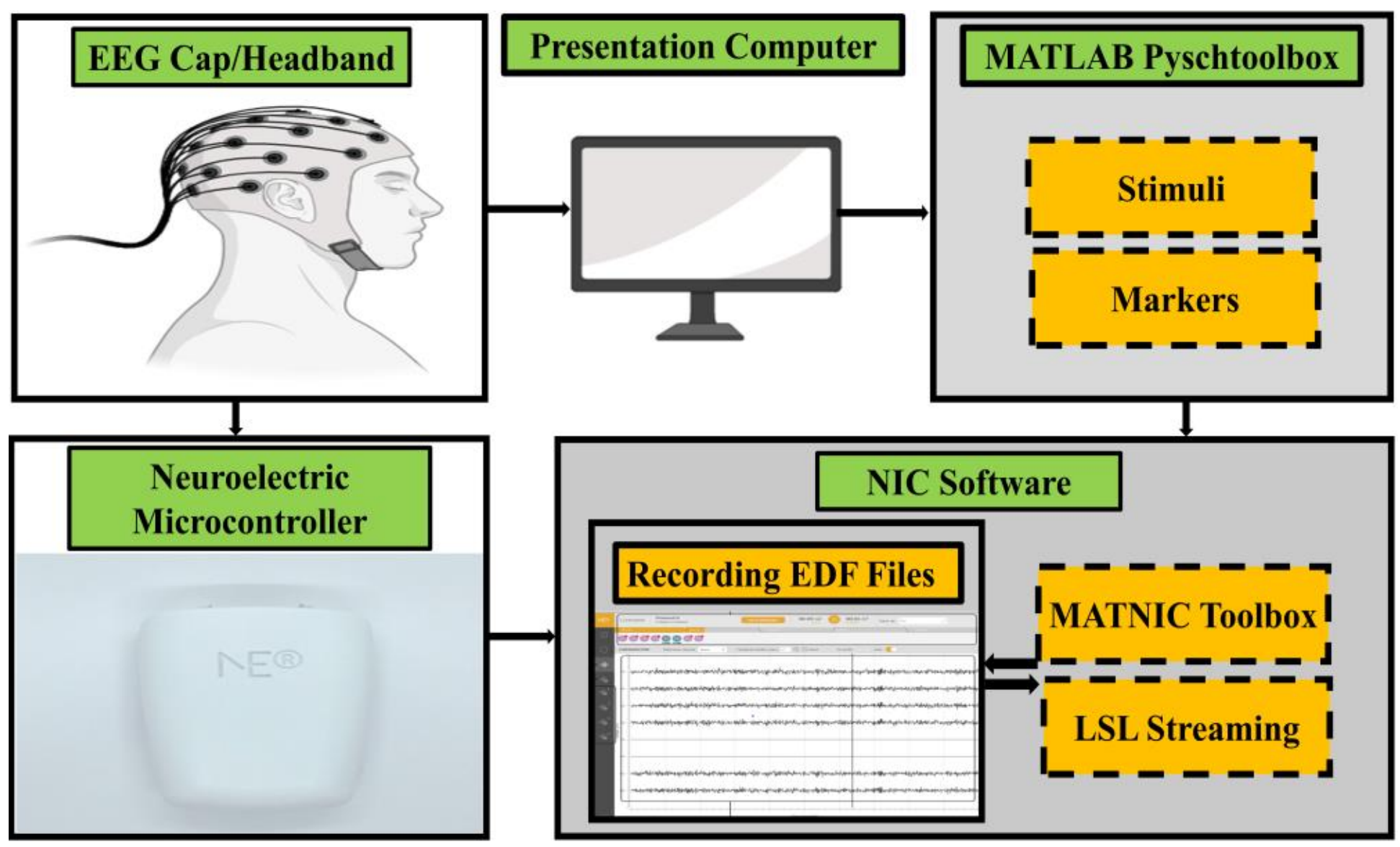

**Fig. 3.** Workflow of EEG signal acquisition using the smart headband with embroidered electrodes, Enobio controller, and NIC software.

During the experimental session, electrical signals were detected on the participant's scalp using the embroidered electrode-embedded smart headband. Embroidered electrodes were placed on the forehead, where little to no hair is present, to avoid the need for skin preparation [37]. Hair increases contact impedance, which can degrade signal quality. The embroidered electrodes were connected to the Enobio microcontroller via snap fasteners [38], [39], [40]. This microcontroller was fixed with a velcro strip to the headband centrally at the back of the participant's head for EEG recording and wirelessly linked to the Neuroelectrics® Instrument Controller (NIC) software to record the EEG data. The controller's light weight (89mm x 61.21mm x 23.8mm and ~58g) ensured that it caused no discomfort to participants during EEG measurement [40]. Once the electrodes were correctly positioned, the functionality of the smart headband was verified by visualising an EEG signal from the customised NIC software. Subsequently, participants were seated in a dimly lit room on a comfortable chair, positioned 60 cm from a computer screen. They were instructed to watch the screen for the presentation and to follow the instructions related to the experimental EEG recording tasks. Details of the behavioral tasks are provided in Section III. B.

Stimuli and instructions related to the behavioral tasks were presented on a display screen configured using Psychtoolbox 3.0.12 in MATLAB (the mathWorks Inc., Natick, Massachusetts, USA). An additional MATNIC toolbox was employed to link the NIC software of the Enobio microcontroller with MATLAB's Psychtoolbox through the Lab Streaming Layer (LSL) [41]. This MATNIC toolbox synchronises the LSL stimuli presentation with event markers associated with the presentation stimuli and transmits these markers to the EEG recording within the NIC software. The raw EEG data were then recorded via the Enobio microcontroller and wirelessly transmitted to customised NIC software (see Fig. 3).

Fitting the smart headband, including the placement of the modular embedded electrodes, took approximately 5 minutes, and performing the behavioral experimental tasks for recording EEG took 10–15 minutes; in total, the experiment took approximately 15–20 minutes. Following this, a commercial EEG system (Brain Products, Munich, Germany), property of the Center for Mind/Brain Sciences (CIMeC) [42], was used for cross-comparative evaluation of the smart headband EEG measurement. This commercial EEG system consisted of 64 sponge-based electrodes, which were soaked in a saline (physiological) solution before contacting the scalp, which is neutral and safe for delicate skin. The commercial EEG setup required approximately 10–15 minutes for scalp preparation, electrode placement, and impedance adjustment.

The screen presentation protocols were identical to those used with the smart headband; however, the EEG signals were recorded using the Brain Vision Recorder software, which is connected to the commercial EEG system. Once preparation was complete, the same experimental series of task-related protocols started, identical to those in the first phase. Following that, the completion of the task-related presentation (lasting approximately 10–15 minutes), participants were asked to remove the headsets. The total duration for each participant using the commercial EEG system was approximately 20–30 minutes, excluding any break times. At the end of the experimentation, the commercial cap was removed, and participants were offered the opportunity to wash their heads in a bathroom adjacent to the experimental room. The sponge-based EEG electrode system was moistened with physiological solution and rigorously sterilised after each participant's session.

## III. Validation of Smart Headband Design

Following ethical approval and recruitment of the participants, validation of the smart headband design involved: the behavioral experimental design task; an EEG analysis; and administration of a survey examining EEG measurement for comfort level.

### *A. Ethical Approval and Recruitment of Participants*

In order to test the performance of the smart headband, this study was conducted at the Center for Mind/Brain Sciences (CIMeC), University of Trento, Italy, and involved both undergraduate and postgraduate students. Prior to starting the study, ethical approval was sought and granted from the Southeast Technological University, Waterford, Ireland, and the University of Trento, Italy. Participants willing to participate received an information leaflet informing them of what the study entailed. Ten participants, including six females and 4 males, aged 32.2 ± 10 years, provided written consent.

*B. Experimental protocol*

Participants wore smart headbands, which was connected to the EEG measurement system via the controller and NIC software, and were asked to sit comfortably while the setup was prepared. Then the experimental protocol started (Fig. 4):

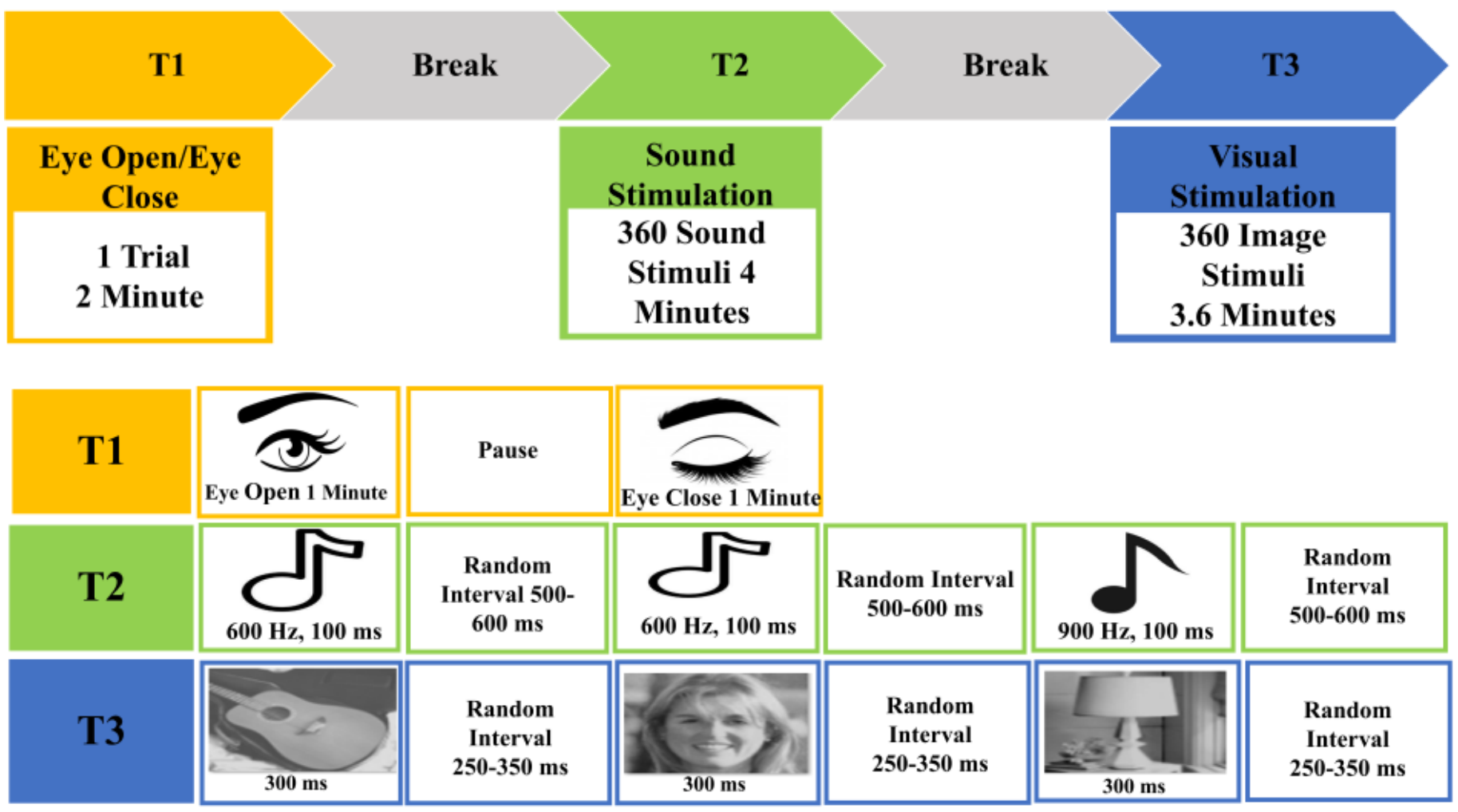

**Fig. 4.** Tasks (T1-T4) and their sequence within the experiment

Task protocol 1 (eyes open/closed): Participants were instructed to sit in a chair in a resting or relaxed state with their eyes open for one minute. Following that, participants were asked to close their eyes for 1 minute.

Task protocol 2 (auditory oddball): Participants listened to two different types of sounds within an oddball paradigm [35]. The standard stimulus was a 600 Hz tone presented 80 % of trials, while the deviant stimulus was a 900Hz Sinewave presented 20% of trials. Each auditory stimulus lasted for 100 milliseconds, with the inter-stimulus interval (ISI) randomly set between 500 and 600 milliseconds. A total of 360 trials were presented in a randomised order, divided into four blocks of 90 trials each. Breaks were provided between blocks to ensure the participant's comfort, and the next block started when the participant was ready. During each block, in order to push them to keep their attention, participants were asked to count the number of deviant stimuli they heard. This task lasted approximately 3–4 minutes, excluding breaks.

Task protocol 3 (visual oddball): Participants were shown a series of images (the same used in the study [43]). Fast periodic presentation of natural face images reveals a robust face-selective electrophysiological response in the human brain [44]. A total of 360 images (trials) were presented, divided into four blocks of 90 images each. Two types of stimuli were used: standard stimuli (e.g., objects) and deviant stimuli (e.g., faces). The image sequence was block-randomised, with each image displayed for 300 milliseconds. The inter-stimulus interval (ISI) was randomly set between 250 and 350 milliseconds. The total duration of this task was approximately 3–4 minutes, excluding breaks. Throughout the task, in order to push them to keep their attention, participants were instructed to count the number of deviant (face) stimuli presented in each block. As in the auditory task, breaks were provided between blocks to maintain participant comfort.

### *C. EEG Data Analysis*

We processed the raw EEG signals recorded during the experimental tasks using both the smart headband and the commercial EEG system with the very same pipeline of data analysis.

#### *1) Pre-processing Pipeline*

The raw EEG data were imported into the MATLAB EEGLAB software [45] and bandpass filtered between 0.3 and 30 Hz with the default EEGLAB filter to remove DC and high-frequency noise [46]. The filtered EEG data were visually inspected, and intervals containing nonstereotyped paroxysmal artifacts were discarded. Bad channels were identified with the LOF algorithm [46]. For the smart headband EEG, zero to maximum one channel was discarded, while for the commercial cap, on average 2-3 channels per participant were discarded. To identify and remove stereotypical artifacts, the default EEGLAB Independent Component Analysis (ICA) decomposition was computed on the concatenation of all segments [46]. Blinks, eye movements, and other topographically localized artifacts were discarded by removing the corresponding independent components identified by visual inspection of their topography and spectro-temporal profile, supported by the classification provided by ICLABELS [48]. EEG signals in bad channels were interpolated with the EEG signals from neighboring channels (standard spherical interpolation method in EEGLAB).

To properly compare the task-related EEG responses measured with the two systems, the commercial EEG data was re-referenced to the electrode channel (Fz), which was employed as a reference channel for the headband EEG data. The EEG recorded data was then divided into three portions corresponding to the segments associated with each task (T1-T3). For a fair comparison in our final analysis, we selected only the electrodes from the commercial EEG cap that correspond to the locations of the electrodes in the smart headband.

#### *2) Data Analysis*

The two EEG segments corresponding to the "eyes opened" and "eyes closed" conditions (linked to Task 1) were designated for the analysis of alpha waves. These selected EEG segments were further inspected to ensure that any remaining eye blinks or noisy intervals were identified and eliminated [35]. Following that, the power spectrum was applied to analyses the power spectral density that indicates the outcomes associated with the eye-opened and eye-closed conditions [14].

Tasks 2 and 3, associated with the classical auditory and visual oddball paradigm, were employed to assess cognitive performance through event-related potential (ERP) [49]. The steady-state evoked potentials were incorporated by presenting 80% of standard stimuli and 20% of deviant stimuli (target) in random order [50]. In order to evaluate the ERP-related tasks (2 and 3), further post-processing steps were undertaken: (1) data were epoched time-locked to the onset of the stimuli (time range: [-0.1 0.5] s); (2) the baseline correction in the interval [-0.1 0] s was applied; and (3) an automatic epoch rejection threshold of 100 μV was established to reject epochs containing residual excessive noise or movement artefacts.

#### *3) Statistical Analysis*

We tested the statistical significance of differences in the power spectrum within the alpha range (0–30 Hz) for the eyes-closed versus eyes-open conditions, and between ERPs in the standard and oddball conditions within the latency range of –100 to 500 ms, using the non-parametric cluster-based test [51] implemented in Fieldtrip [52]. This method allows statistical testing with no need of a priori selection of spatial ROIs because it controls for multiple comparisons by clustering neighboring channel pairs that exhibit statistically significant effects (test used at each channel point: dependent-samples t statistics, threshold: p=0.05) and using a permutation test to evaluate the statistical significance at the cluster level (Montecarlo method, 1000 permutations for each test).

### *D. Survey on the comfort level of each EEG system*

Following the EEG signal measurement during the experimental tasks session, participants were invited to complete a survey about their comfort with the smart headband compared to the commercial EEG system. These closed-ended questionnaires included: firstly, to rank the overall comfort level of wearing the smart headband compared to the commercial EEG cap, with 1 indicating the lowest and 5 indicating the highest; secondly, to rate any discomfort or skin irritation experienced from wearing the embroidered electrodes on their forehead or when using the commercial EEG cap. The scale went from 1 (no discomfort) to 5 (highest amount of discomfort or irritation); thirdly, to describe the sensation they experienced on their forehead while wearing the smart headband with the microcontroller attached; and finally, if they felt stable and secure while wearing a smart headband on their foreheads.

## IV. RESULTS

### *A. Tasks 1 (eyes closed/eyes open)*

Task 1 was designed to test whether we could reliably record with the smart headband an increase in the Alpha frequency band (8-12 Hz) from eyes open to eyes closed. We computed the power spectrum of EEG signal measurements obtained from our smart headband and the conventional sponge-based commercial EEG cap separately during the eyes-closed and eyes-open sessions (Fig. 5).

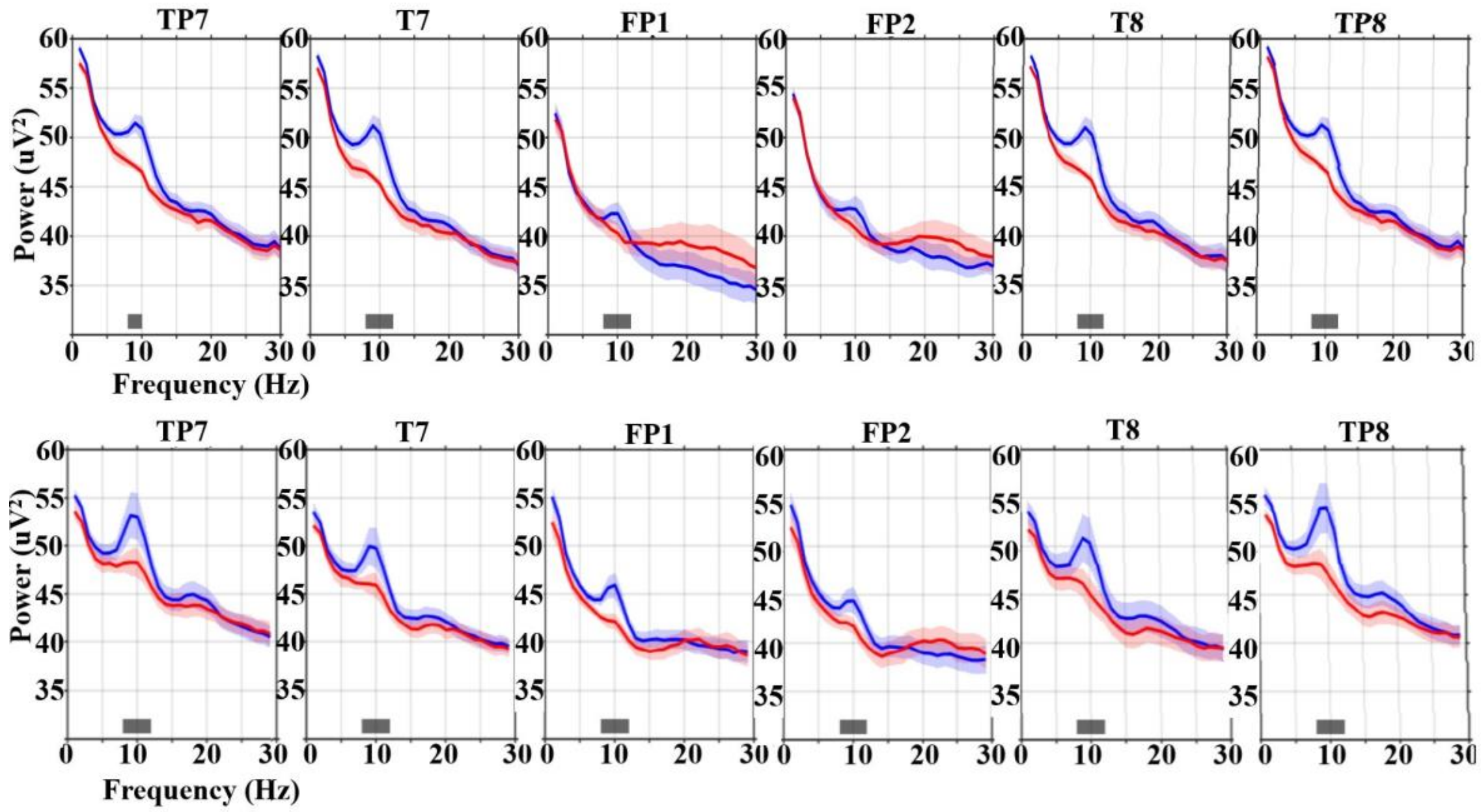

**Fig. 5.** Task 1. Both EEG systems detect the Alpha-band increase from eyes-open to eyes-closed. X-axis – Frequency in (Hz), Y-axis – Power Spectral Density in ($\mu V^2$). The first row EEG measurement is from our smart headband, while the second row is from a commercial sponge-based headset. The blue and red line graphs illustrate eye-closed and eye-open conditions, respectively. The light-colored shaded area displays the standard deviation of the mean of the alpha wave activity from 10 individual participants, with the thicker line indicating the mean value. The horizontal black bar illustrates the significant latency values for each respective electrode channel.

Both devices presented robust Alpha wave activity (8–12 Hz) during eye closure. The mean peak amplitudes of the Alpha waves were comparable across both the devices and the electrodes, in the range 45 – 50 $\mu V^2$ for our smart headband and 45 – 55 $\mu V^2$ for the commercial cap, indicating comparable signal acquisition performance. Electrodes TP7, T7, T8, and TP8 had the strongest alpha peak, showing the most significant effect in both the smart headband and the commercial cap. The cluster-based statistical analysis of the difference between the Alpha-band power in the eyes closed and, in the eyes, open condition (see Methods) identified a positive cluster in both systems (smart headband p-value: 0.015;  commercial cap p-value: 0.044), confirming that the smart headband detects the alpha-band increase even more significantly than the commercial EEG cap. The results were consistent with literature findings [9], [35], [53], with eye blink analysis, particularly useful for drowsiness detection and other neurofeedback applications [14].

### *B. Task 2 (Auditory Oddball)*

The smart headband reliably detected auditory ERPs at comparable latencies to the sponge-based commercial cap. To test the auditory oddball effect, we computed the P300 ERP wave, which should be present for deviant stimuli but not for standard ones [54].

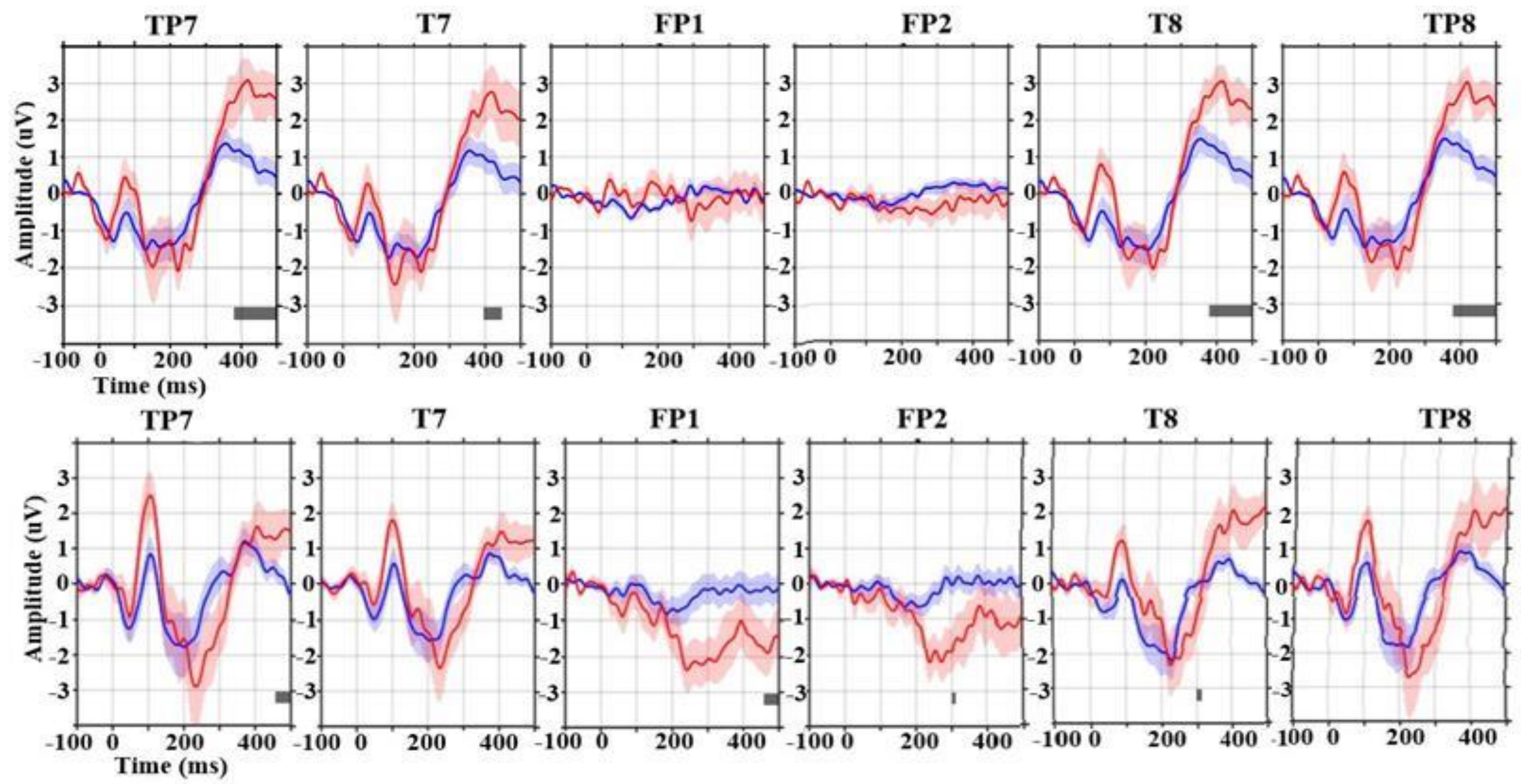

**Fig. 6.** Task 3 Both EEG systems detect the Auditory Oddball effect. X-axis – Time in (ms), Y-axis – Amplitude in ($\mu V^2$). The first row is the ERP response from our smart headband, while the second row is from a commercial sponge-based headset. The blue and red line graphs illustrate the ERPs from the standard and deviant stimuli responses, respectively. The light-colored shaded area represents the standard deviation of the mean of the responses from individual participants, while the thicker line graph represents the mean values. The horizontal black bar illustrates the significant latency values for each respective electrode channel.

Fig. 6 represents the auditory paradigm ERP generated from the smart headband and commercial cap. The amplitude of both systems is almost similar, ranging from approximately -4 to +4 $\mu V^2$. While the ERPs from the smart headband are slightly noisier compared to the commercial cap, both systems clearly show an initial wave at around 100 ms and a clear P300 starting at 250 ms and peaking at 450–500 ms after stimulus onset. For the smart headband, the P300 difference between deviant and standard stimuli was statistically significant in a bilateral cluster including electrodes TP7, T7, T8, and TP8 ($p = 0.014$, latency: 370–500 ms). The commercial cap also displayed an effect in the expected time window ($p = 0.064$, 300–350 ms), though this did not reach the threshold for statistical significance. Notably, while the P300 is typically more visible from central electrodes [55]. The outer hair area electrode channels are still able to detect significant amplitude changes within the P300 response range across the majority of channels.

*C. Task 3 (Visual oddball)*

In the visual paradigm, the participant's brain activity is reflected in the target response in the form of a negative N170 component, representing the early attention sensory processes, and the activity of the parietal stream in the visual cortex, followed by the P300, which signifies consciousness or cognitive evaluation [56]. Both devices presented a robust negative peak between 150 – 200 ms, corresponding to the N170, a face-sensitive ERP component, typically maximal at occipital-temporal sites such as TP7, T7, T8, and TP8, and is reported to originate from lateral temporal regions behind the ear. Interestingly, in our recordings, the N170 was more pronounced than the P300, which is usually less visible at temporal sites (see Fig. 7).

The P300 is generally strongest over centro-parietal regions, particularly at midline sites such as Cz. However, due to the design of our smart headband, which places electrodes laterally outside the hair for simplicity, we were unable to cover central-parietal midline regions. As a result, the P300 response appears weaker, whereas the N170 response is more prominent. This observation is consistent with the electrode coverage of both devices. For the N170 effect, significant differences between standard and deviant cluster responses were most evident at the temporal electrode cluster, including TP7, T7, T8, and TP8 ($p = 0.013$, latency: 150–200 ms). The commercial EEG cap also shows a significant effect within this time window ($p < 0.001$, latency: 150–200 ms), confirming a robust difference between standard and deviant responses in both systems.

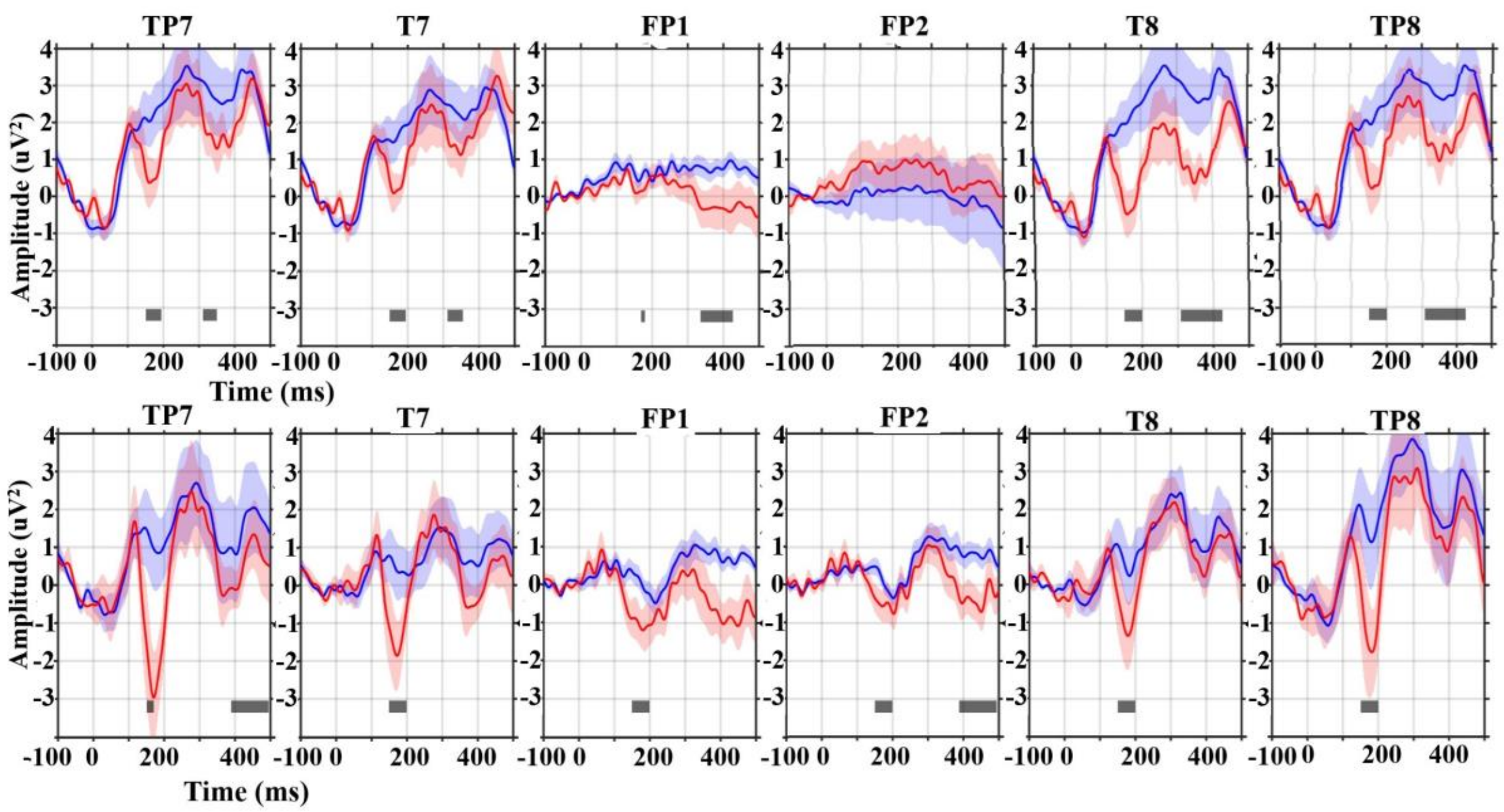

**Fig.7.** Both EEG systems detect the Visual Oddball effect**.** X-axis – Time in (ms), Y-axis – Amplitude in ($\mu V^2$). The first row is the ERP response from our smart headband, while the second row is from a commercial sponge-based headset, recorded during the same task. The blue and red line graphs illustrate the ERPs from the standard and deviant stimuli responses, respectively. The light-colored shaded area represents the standard deviation of the mean of the responses from individual participants, while the thicker line graph represents the mean values. The horizontal black bar indicates the time window of significant latency values for each respective electrode channel.

*D. Assessment of comfort*

The survey assessing the comfort level of the smart headband compared with the commercial EEG cap is presented in Fig. 8. The subjects were asked to rate their global comfort level regarding the EEG setup, which includes aspects such as wearing the EEG device, preparation time, carrying the controller at the back of the head, and the environment during EEG recording. Out of ten subjects, six rated their comfort level as high (5), while two selected (4), indicating a high level of comfort. Only one subject rated their comfort level as 3, reflecting a medium level of comfort with the smart headband, and no one selected the lowest or minimum level of comfort (see Fig. 8(a)). In contrast, the same question was posed regarding the commercial cap presented in Fig. 8(b). Three subjects rated their comfort level as high (5), two selected (4), and two subjects chose a medium level of comfort (3). However, three subjects perceived the commercial cap as having the lowest level of comfort by selecting (2). Fig. 8 (c) presents the findings related to the questions posed regarding discomfort or skin irritation. For the smart headband, 9 participants reported no discomfort or skin irritation (1) while wearing the embroidered electrode-integrated smart headband. However, the commercial cap illustrated in Fig. 8(d), in which 6 participants rated their experience as high (1), indicating no skin irritation or discomfort. One participant reported very minimal discomfort, while three participants indicated a medium level of skin irritation or discomfort (3) while wearing the commercial cap.

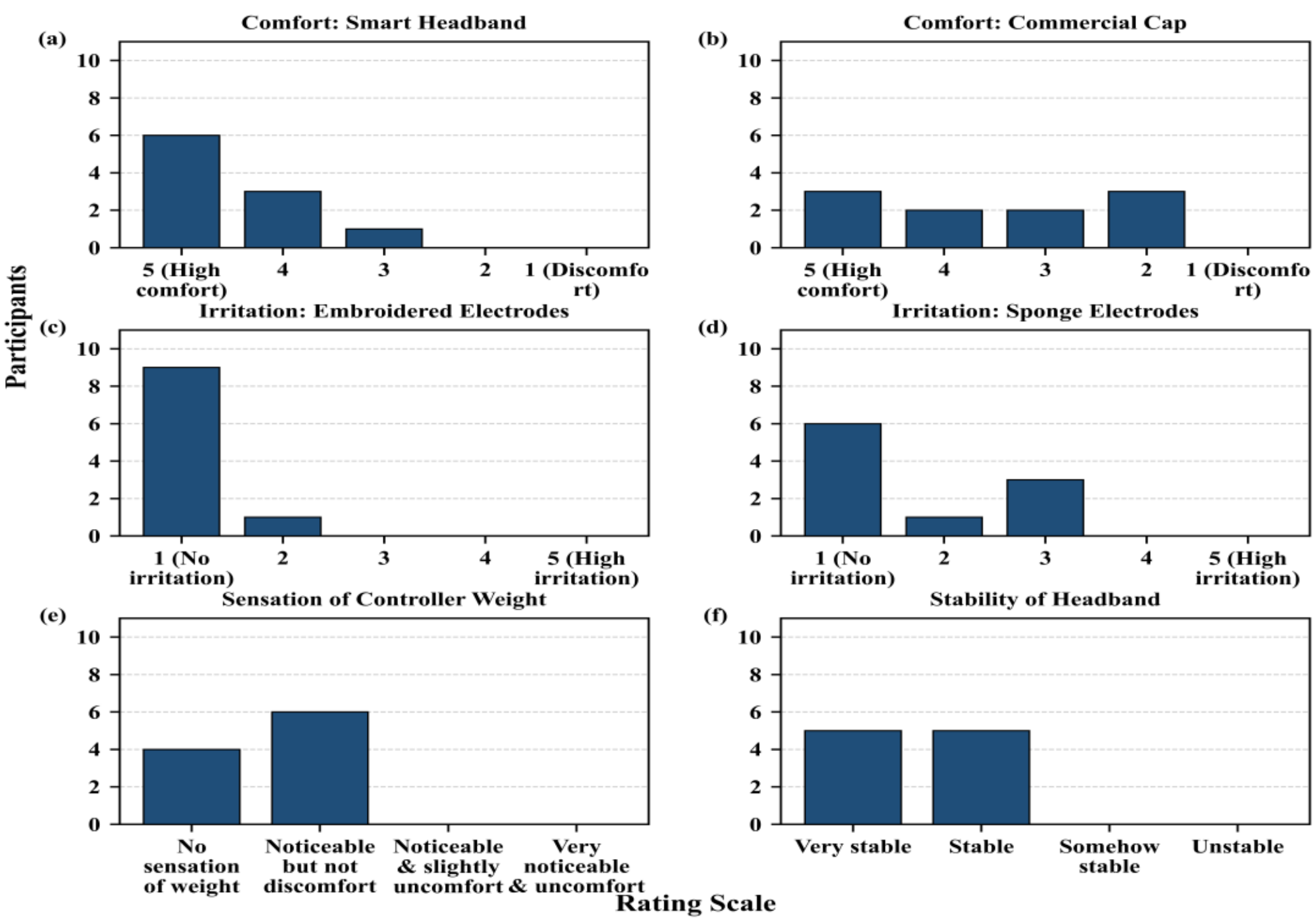

**Fig. 8** Survey results in relation to the comfort level of both EEG setups.

Survey results consistently favored the smart headband, with 6 out of the 10 participants rating its comfort as 'high'. The findings provided a negative indication of discomfort or irritation in wearing the smart headband, with the participants reporting positive feedback for secure wearability and stability of the smart headband, with only a mild cognizance of the microcontroller unit

## V. DISCUSSION

This study designed and developed a non-invasive modular smart headband with embroidered electrodes for EEG measurement. Its performance was evaluated through human trials and compared with a commercial sponge-based EEG cap, through comparison of captured EEG potentials: (1) alpha wave rhythms (eye-blinks), (2) P300 auditory responses, and (3) N170 visual responses. The designed smart headband was able to acquire and reproduce the morphology, latency, and the ERP-specific signatures associated with each of the tasks. Our findings demonstrate that alpha waves and visual oddballs show almost similar latency ranges in both the smart headband and the commercial EEG cap. However, the smart headband captures even better latency than the commercial EEG for the auditory oddball effect. This effect is visible from the bar plots presented in the following figures. Through behavioral experimental tests, this paper establishes a paradigm study showing embroidered electrodes as a viable option to realize continuous non-invasive brain monitoring.

The smart headband was designed to enhance the ease of wearing and comfort while reducing device complexity through minimal use of electrodes. The embroidered electrodes are soft and flexible, eliminating the need for electrolytic gel, which can cause skin irritation [2], [12], [29]. By eliminating the need for disposable gel electrodes and cumbersome wiring, the designed smart headband offers a reusable, adjustable, and more comfortable solution for long-term monitoring [9]. The findings of the survey relating to comfort level suggested that this prototype design of a smart headband can be wearable for long-term brain signal monitoring. Additionally, these electrodes reduce the costs associated with disposable gel-based medical-grade electrodes, as they are reusable and can maintain hygienic compliance through multiple washes.

The current literature shows a complex multi-channel montage EEG utilized for the maximization of accuracy in laboratory and clinical studies [59]. However, practical usability studies presented that end-users prefer EEG systems that are practical to wear with minimal interfaces and wires [35]. This study utilised a limited number of electrodes to perform stipulated measurements. However, when utilizing fewer electrode montages, task-relevant placement of electrodes significantly impacted the acquired signals to analyse the said behavioral tasks [35]. Nevertheless, encouraging evidence was found even with this limited montage associated with auditory and visual stimuli.

***Limitations in the study:*** The fabrication of embroidered electrodes is straightforward, easy to adjust, cost-effective, and time efficient. Nevertheless, the pilot study presented comes with its limitations. First, only a small number of participants (n=10) were tested converges on the statistical analyses, thereby restricting generalizability. In the future, a larger number of participants will be required to validate the findings on a broader scale. Second, the use of the smart headband for non-critical applications could be validated at this stage, but the reduced number of montages restricts its suitability for clinical diagnostics. Third, due to the intensive nature of laboratory experimentation, this study included only closed-ended questions related to comfort. However, in the future, incorporating open-ended questions would provide an opportunity to gain deeper insights into participants' views on the smart headband design and its potential improvements. Finally, from a hardware perspective, the current setup utilized a Neuroelectrics microcontroller. Future iterations could employ an Arduino board to reduce the cost of the hardware setup further.

## VI. Conclusion and Future Work

In this study, a textile-based embroidered electrode-integrated smart headband was designed, fabricated, and implemented. The application of this smart headband is straightforward and does not require any skin preparation. The super-soft and flexible design of the embroidered electrode ensures comfort for long-term EEG biopotential recording. Furthermore, the modular design of the embroidered electrode allows for adjustment according to the 10-20 electrode placements, and the electrodes can be interchanged or removed for washing the headband, thereby maintaining hygienic compliance.

The results illustrate that the neurocognitive responses detected from the smart headband are equivalent to those from the commercial EEG setup. However, the limited number of embroidered electrodes in the integrated headband may restrict the amount of information obtained compared to the commercial cap. Nevertheless, this initial pilot study serves as proof of concept for EEG measurement outside of clinical and laboratory settings. For future research, increasing the number of electrodes or incorporating multimodal input sensing may offer improved accuracy for applications related to neurological disorders, such as epilepsy, parkinson's disease, unconscious state, sleep disorders, and autism-related conditions.

Future research will explore the integration of additional textrodes into the smart headband in a modular and interchangeable headband design method, enabling an enhanced level of EEG measurements through a larger dataset of EEG measurements. Additionally, further focus on the integration and utilisation of embroidered electrodes, in order to obtain multiple physiological signal measurements such as EEG and EMG, will be carried out, which will lead to the creation of a wearable smart hybrid garment offering an unobtrusive wearable solution for health-related applications.

## GENERATIVE AI DECLARATION

Grammarly was used for spelling and grammar checks.